\documentclass[twocolumn,showpacs,preprintnumbers,amsmath,amssymb]{revtex4}

\usepackage{graphicx}
\usepackage{dcolumn}
\usepackage{bm}

\begin{document}


\title{The Dirac Conjecture and the Non-uniqueness of Lagrangian}
\author{Yong-Long Wang$^{1,2,3}$}
 \email{wylong322@163.com}
\author{Chang-Tan Xu$^{2}$}%
\author{Hua Jiang$^{2}$}
 \email{jianghua@lyu.edu.cn}
\author{Wei-Tao Lu$^{2}$}
\author{Hong-Zhe Pan$^{2}$}
\author{Hong-Shi Zong$^{4,5,6}$}
\email{zonghs@chenwang.nju.edu.cn}
\address{$^{1}$ Key Laboratory of Modern Acoustics, MOE, Institute of Acoustics, and Department of Physics, Nanjing University, Nanjing 210093, P. R. China}
\address{$^{2}$ Department of Physics, School of Science, Linyi University, Linyi 276005, P. R. China}
\address{$^{3}$ Center for Theoretical Physics, Massachusetts Institute of
Technology, Cambridge, MA02139, USA}
\address{$^{4}$ Department of Physics, Nanjing University, Nanjing 210093, P. R. China}
\address{$^{5}$ Joint Center for Particle, Nuclear Physics and Cosmology, Nanjing 210093, P. R. China}
\address{$^{6}$ State Key Laboratory of Theoretical Physics, Institute of Theoretical Physics, CAS, Beijing 100190, China}

\date{\today}

\begin{abstract}
By adding the total time derivatives of all the constraints to the
Lagrangian step by step, we achieve the further work of the Dirac
conjecture left by Dirac. Hitherto, the Dirac conjecture is proved
completely. It is worth noticing that the addition of the total
time derivatives to the Lagrangian can turn up some constraints
hiding in the original Lagrangian. For a constrained system, the
extended Hamiltonian $H_E$ considers more constraints, and shows
symmetries more obviously than the total Hamiltonian $H_T$. In the
Lagrangian formalism, we reconsider the Cawley counterexample, and
offer an example in which in accordance with its original
Lagrangian its extended Hamiltonian is better than its total
Hamiltonian.
\end{abstract}

\maketitle

In order to quantize systems with a singular Lagrangian, Dirac
proposed that all first-class constraints are generators of gauge
transformations, and the Hamiltonian should contain all first-class
constraints by Lagrange multipliers as the extended Hamiltonian which is
denoted by $H_E$\cite{1}. According to $H_E$, Dirac offered a
canonical procedure for modern quantum field theory\cite{2},
which has been developed widely and deeply. Summarizing the results
over half a century, it can be said that the relevance of
the Dirac procedure paves the way to the Faddeev-Senjanovic path
integral quantization approach\cite{3,4}, helps to find a complete
set of constraints for the Faddeev-Jackiw quantization
formalism\cite{5,6}, and furnishes the classical basis for the
powerful Becchi-Rouet-Stora-Tyutin-Batalin-Fradkin-Vilkovisky
(BRST-BFV) gauge field quantization procedure\cite{7,8,9,10,11}.

At present, most points about the Dirac procedure have been well
understood. However, the Dirac conjecture is still not completely
proved. Dirac has proved that all the primary first-class
constraints $\phi_a$, and the Poisson brackets
$[\phi_a,\phi_{a^{\prime}}]$ of two arbitrary primary first-class
constraints are the generators of gauge transformations\cite{2}.
He left us a further work, i.e., the proof that the Poisson
brackets $[\phi_a,H^{\prime}]$ of the first-class Hamiltonian
$H^{\prime}$ with an arbitrary primary first-class constraint are
the generators of gauge transformations\cite{12}. Here we note
that in the above Poisson brackets, $\phi_a$ just denotes a
primary first-class constraint. In order to prove the Dirac
conjecture, on the basis of the original primary first-class
constraints, Castellani et al. redefined new independent
first-class constraints\cite{13,14}, which generate higher-stage
constraints in closed form but without cross term\cite{15}. With
the reconstructed constraints, the Dirac conjecture is proved
easily, and the number of the independent generators of gauge
transformations is obviously limited by that of the primary
first-class constraints\cite{13,14,15}. However, this
reconstruction is of no great help not only practically, since the
necessary redefinitions may be quite complicated, but also
theoretically\cite{16}. Henneaux and Teitelboim then proved the
Dirac conjecture under four restricted conditions, which are: no
mixture among first- and second-class constraints, no bifurcations
in the procedure of consistency algebra, the functions $V_a^b$
appearing in $[H,\phi_a]=V_a^b\phi_b$ and obeying appropriate rank
conditions on the constraint surface, and the first-class
constraints being irreducible, respectively\cite{17}. Absorbing
all primary second-class constraints into the canonical
Hamiltonian $H_C$, we obtain a Halmitonian $H^{\prime}$ which is
first-class. Using the Dirac-Bergmann method, Batlle and coauthors
tried to find a complete set of constraints for a constrained
system\cite{18}. Using this method, Cabo tried to prove the
validity of the Dirac conjecture\cite{19}, but he might lose some
first-class constraints generated by the consistencies of
second-class constraints. All the previous discussions are in the
Hamiltonian formalism. In the Lagrangian formalism, Lusanna
extended the second Noether theorem, and used it to discuss the
Dirac conjecture and obtain an affermative
answer\cite{20,21,22,23}.  Recently, by introducing auxiliary
variables the original Lagrangian is replaced by an extended
Lagrangian including all higher-stage (secondary, tertiary,
$\cdots$) constraints, from which the Dirac conjecture is
satisfied automatically \cite{24,25}.

In this paper we will directly prove the Dirac conjecture without any
restriction, by adding the total time derivatives of all the
constraints to the Lagrangian step by step. Our procedure covers
all first-class constraints, irrespective of whether they are deduced from the
consistencies of first-class constraints or second-class ones, or
whether they are generated by primary constraints or higher-stage
ones. In our procedure the higher-stage first-class constraints
play the same role as the primary first-class ones, because the
addition of the total time derivatives of all constraints ensure
that all higher-stage constraints appear in its new Lagrangian and
its new total Hamiltonian. This method should of course belong to the
Lagrangian formalism.

Following Dirac, we consider a constrained system with the Lagrangian
$L(q_i,\dot{q}_i) (i=1,\cdots,N)$, in which there are primary
constraints
\begin{equation}
 \phi_{m}(q,p)\approx 0 \quad (m=1,\cdots, M),
 \end{equation}
and higher-stage constraints
 \begin{equation}
\chi_{k}(q,p)\approx 0 \quad (k=1,\cdots,K),
\end{equation}
where $M=N-R$ is determined by the rank $R$ of the matrix
$\frac{\partial^2 L}{\partial {\dot{q}_i\partial {\dot{q}_j}}}$,
and $K$ is the number of the higher-stage constraints. They are
collectively denoted by
\begin{equation}
\phi_{j}\approx 0 \quad (j=1,\cdots, M+K).
\end{equation}
The corresponding total Hamiltonian $H_T$ is
\begin{equation}
H_T=H_C+U_{\tilde{m}}\phi_{\tilde{m}}+\mu_{m^{\prime}}\phi_{m^{\prime}}=H^{\prime}+\mu_{m^{\prime}}\phi_{m^{\prime}},
\end{equation}
where $H_C=p_i\dot{q}_i-L(q_i,\dot{q}_i)$, $\mu_{m^{\prime}}$ is a
parameter corresponding to the first-class primary constraint
$\phi_{m^{\prime}}$ which is an arbitrary function of only time, and
$U_{\tilde{m}}$ is a function of only the $q$'s and the $p$'s which is
determined by
\begin{equation}
U_{\tilde{m}}=-[\phi_{\tilde{m}},\phi_{\tilde{n}}]^{-1}[\phi_{\tilde{n}},H],
\end{equation}
where $\tilde{m},\tilde{n}=1,\dots,\tilde{M}$ with $\tilde{M}$
being the number of the second-class primary constraints,
$m^{\prime}=1,\dots,M^{\prime}$ with $M^{\prime}$ being the number
of the first-class primary constraints, and
$\tilde{M}+M^{\prime}=M$. $\tilde{M}$ is easily determined by the
rank of the matrix $[\phi_m,\phi_n]_{M\times M}$ of the primary
constraints. In this case, there is no more first-class constraint
that can be combined with the $\phi_{\tilde{m}}$'s.

According to $H_T$ (4), for a general dynamical variable $g$
which depends only on the $q$'s and the $p$'s and has initial value
$g_0$, its value at time $\delta t$ is
\begin{equation}
g(\delta t)=g_0+\delta
t\{[g,H^{\prime}]+\mu_{m^{\prime}}[g,\phi_{m^{\prime}}]\}.
\end{equation}
Owing to that $\mu_{m^{\prime}}$ is arbitrary and $\delta
t$ is small, $\phi_{m^{\prime}}$ is a generator of gauge
transformations. The Poisson brackets
$[\phi_{m^{\prime}},\phi_{n^{\prime}}]$ of two arbitrary primary
first-class constraints are also generators of gauge
transformations, which is proved by the subtraction between the
result of applying two contact transformations with generating
functions $\mu_{m^{\prime}}\phi_{m^{\prime}}$ and
$\mu_{n^{\prime}}\phi_{n^{\prime}}$ and that of applying the two
transformations in succession in reverse order, and with
Jacobi's identities.  The difference is
\begin{equation}
\Delta
g=\mu_{m^{\prime}}\mu_{n^{\prime}}[g,[\phi_{m^{\prime}},\phi_{n^{\prime}}]],
\end{equation}
where $\mu_{m^{\prime}}\mu_{n^{\prime}}$ is arbitrary. For the
sake of completeness, Dirac supposed that the Poisson brackets
$[\phi_{m^{\prime}},H^{\prime}]$ are also generators of gauge
transformations, because both $H^{\prime}$ and $\phi_{m^{\prime}}$
are first-class. After about forty years, this supposition was
proved by Henneaux and Teitelboim with some restrictions\cite{17}.
If the constraint $\phi_m$ is second-class but the Poisson bracket
$[\phi_{m},H^{\prime}]$ is first-class, whether the
$[\phi_{m},H^{\prime}]$ is generator of gauge transformations is
still left to be undertermined.

It is well known that the addition of a total time derivative or a
total space-time derivative to a Lagrangian does not change its
equations of motion\cite{26,27}, which is called the
non-uniqueness of the Lagrangian. Thus, we could go over to a new
Lagrangian
\begin{equation}
L^{1}(q,\dot{q})=L(q,\dot{q})-\frac{d(\mu_{m}\phi_{m})}{dt},
\end{equation}
and the $H_T$ (4) is replaced by
\begin{equation}
H^{1}_T=H^{\prime}+\mu_{m^{\prime}}\phi_{m^{\prime}}+\frac{d(\mu_{m}\phi_{m})}{dt},
\end{equation}
where the $H^{\prime}$, $\mu_{m^{\prime}}$ and $\phi_{m^{\prime}}$
are the same as the corresponding ones in (4), the third term in
r.h.s. of (9) is determined by the $H_T$ (4), and $\mu_m$ is an
arbitrary function of time, which is completely arbitrary when
$\frac{d}{dt}\phi_m$ is first-class, and is completely determined
when $\frac{d}{dt}\phi_m$ is second-class. For the system we
discuss, because of the non-uniqueness of Lagrangian, we know that
$L^{1}$ is just as good as $L$, and $H^{1}_T$ is just as good as
$H_T$. This is the $1$st-stage.

On the basis of the $H^{1}_T$ (9), we can rewrite (6) as
\begin{equation}
\begin{split}
g(\delta t)&=g_0+([g,H^{\prime}+\dot{\mu}_{\tilde{m}}\phi_{\tilde{m}}]+\xi_{m^{\prime}}[g,\phi_{m^{\prime}}]\\
&\quad+\mu_{m}[g,[\phi_{m},H_T]])\delta t,
\end{split}
\end{equation}
where the coefficient
$\xi_{m^{\prime}}=\mu_{m^{\prime}}+\dot{\mu}_{m^{\prime}}$ is
arbitrary because $\mu_{m^{\prime}}$ is arbitrary, and
$\phi_{m^{\prime}}$ is the generator of gauge transformations
because $\delta t$ is small and $\xi_{m^{\prime}}$ is arbitrary.
The $[\phi_{m},H_T]$ in (10) can be expanded as
\begin{equation}
[\phi_{m},H_T]=[\phi_{m},H^{\prime}]+\mu_{m^{\prime}}[\phi_{m},\phi_{m^{\prime}}],
\end{equation}
which can completely generate secondary constraints which we
denote by $\chi_{m_1}$, because $\phi_m$ denotes a primary
constraint (irrespective of whether it is first-class or
second-class). $\chi_{m_1}$ can be classed into second-class
constraints denoted by $\chi_{\tilde{m}_1}$ and first-class ones
denoted by $\chi_{m^{\prime}_1}$. Considering the second-class
secondary constraints $\chi_{\tilde{m}_1}$, the $H^{\prime}$
should be replaced by
$H_1^{\prime}=H^{\prime}+\mu_{\tilde{m}_1}\chi_{\tilde{m}_1}$. The
appearances of the term $\delta t\mu_m[g,[\phi_m,H_T]]$ in (10)
and of the term $[\phi_m,H^{\prime}]$ in (11) achieve the work
left by Dirac, which is that the Poisson brackets
$[\phi_m,H^{\prime}]$ are generators of gauge transformations. It
is obvious that all the secondary first-class constraints
$\chi_{m^{\prime}_1}$ are generators of gauge transformation. We
can generalize the result to higher-stage.

For the sake of simplicity, we use $\theta_{k_1}$ to denote
$(\phi_m,\chi_{m_1})$. They can be divided into first-class
constraints denoted by $\theta_{k^{\prime}_1}$ and
second-class ones denoted by $\theta_{\tilde{k}_1}$. The $H^1_T$
(9) is then rewritten as
\begin{equation}
H^{1}_{T}=H^{\prime}_1+\lambda_{k_1^{\prime}}\theta_{k_1^{\prime}},
\end{equation}
where $H^{\prime}_1=H_C+U_{\tilde{k}_1}\theta_{\tilde{k}_1}$ with
$U_{\tilde{k}_1}=-[\theta_{\tilde{k}_1},\theta_{\tilde{l}_1}]^{-1}[\theta_{\tilde{l}_1},H_C]$,
and $\lambda_{k_1^{\prime}}$ is the Lagrange multiplier
corresponding to $\theta_{k_1^{\prime}}$. We begin the next
annulation to add $-\frac{d(\mu_{m_1}\chi_{m_1})}{dt}$ determined
by the $H^1_T$ (12) to the $L^{1}$ (8) and obtain
\begin{equation}
L^{2}=L^{1}-\frac{d(\mu_{m_1}\chi_{m_1})}{dt},
\end{equation}
and the $H^{1}_{T}$ (12) will be replaced by
\begin{equation}
H^{2}_{T}=H^{\prime}_1+\lambda_{k_1^{\prime}}\theta_{k_1^{\prime}}+\frac{d(\mu_{m_1}\chi_{m_1})}{dt}.
\end{equation}
Note that here the total derivative term in (14) is defined by
$H^1_T$ (12) rather than $H_T$ (4) used by Dirac,
$\frac{d(\mu_{m_1}\chi_{m_1})}{dt}=
\dot{\mu}_{m_1}\chi_{m_1}+\mu_{m_1}[\chi_{m_1}, H^{1}_{T}]$, in
which the Poisson bracket generates tertiary constraints. This is
the $2$nd-stage. Repeating the above process, we can arrive at the
$i$th-stage, in which the total Hamiltonian $H^{i}_{T}$ is
\begin{equation}
H^{i}_T=H_{i-1}^{\prime}+\lambda_{k_{i-1}^{\prime}}\theta_{k_{i-1}^{\prime}}+\frac{d(\mu_{m_{i-1}}\chi_{m_{i-1}})}{dt},
\end{equation}
where the third term in r.h.s. of (15) should be determined by
$H_T^{(i-1)}$ as
\begin{equation}
\frac{d(\mu_{m_{i-1}}\chi_{m_{i-1}})}{dt}=\dot{\mu}_{m_{i-1}}\chi_{m_{i-1}}+\mu_{m_{i-1}}[\chi_{m_{i-1}},H^{i-1}_T].
\end{equation}
It is easy to prove that the Poisson bracket
$[\chi_{m_{i-1}},H^{\prime}_{i-1}]$ is generator of gauge
transformations, when it is a new first-class constraint, because
$\delta t$ is small and $\mu_{m_{i-1}}$ is arbitrary. This
procedure will terminate when the addition of the total time
derivatives of new higher-stage constraints to the Lagrangian does
not generate new constraint. Hitherto, the Dirac conjecture is
proved eventually.

In this procedure, in order to find a complete set of constraints
for a constrained system, we follow the Dirac-Bergmann method
except that in the $i$th-stage the $H_T$ used by Dirac is replaced
by the $H^{i-1}_{T}$. This method can be called the modified
Dirac-Bergmann method. In this method, we consider not only the
constraints generated by the Poisson brackets
$[\theta_{k_i^{\prime}},H^{\prime}]$ of first-class Hamiltonian
$H^{\prime}$ and an arbitrary first-class constraint
$\theta_{k_i^{\prime}}$, and those generated by the Poisson
brackets $[\theta_{k_i^{\prime}},\theta_{k_j^{\prime}}]$ of two
arbitrary first-class constraints\cite{17,18}, but also the ones
generated by the Poisson brackets
$[\theta_{\tilde{k}_i},H^{\prime}]$ of first-class Hamiltonian
$H^{\prime}$ and an arbitrary second-class constraint
$\theta_{\tilde{k}_i}$, and the Poisson brackets
$[\theta_{\tilde{k}_i},\theta_{k^{\prime}_j}]$ of an arbitrary
second-class constraint $\theta_{\tilde{k}_i}$ and an arbitrary
first-class constraint $\theta_{k^{\prime}_j}$.

In the discussion of the Dirac conjecture, some counterexamples
were given. The one given by Cawley has the Lagrangian\cite{28}
\begin{equation}
L=L(x,y,z,\dot{x},\dot{y},\dot{z})=\sum_{n=1}^N(\dot{x}_n\dot{z}_n+\frac{1}{2}y_nz_n^2).
\end{equation}

On the basis of the discussion given by Lusanna\cite{22}, one has
the conjugate momenta $p_{x_n}$, $p_{y_n}$, $p_{z_n}$
\begin{equation}
p_{x_n}=\dot{z}_n,\quad p_{y_n}=0, \quad p_{z_n}=\dot{x}_n,
\end{equation}
and the Euler-Lagrange equations
\begin{equation}
L_{x_n}=-\ddot{z}_n=0, L_{y_n}=\frac{1}{2}z_n^2=0,
L_{z_n}=-\ddot{x}_n+y_nz_n=0.
\end{equation}
Integrating the equations (19) and substituting into the equations
(18), one obtains
\begin{equation}
\begin{split}
& x_n(t)=A+Bt,\quad y_n(t)\quad \text{arbitrary},\quad z_n(t)=0,\\
& p_{x_n}(t)=0,\quad p_{z_n}(t)=B,
\end{split}
\end{equation}
where $A$ and $B$ are integration constants.

Under the Noether transformations $\delta y_n=\epsilon(t)$, one
gets $\delta L=\epsilon(t)D$ with $D=\frac{1}{2}(z_n)^2=0$. There
are the Noether identities
\begin{equation}
p_{y_n}=0,\quad \frac{1}{2}z_n^2=0,
\end{equation}
and the generalized contracted Bianchi identities
\begin{equation}
\sqrt{2L_{y_n}}L_{x_n}-(\frac{d}{dt}\sqrt{2L_{y_n}})^2+\frac{d^2L_{y_n}}{dt^2}=0.
\end{equation}
According to the identities (21), there are $2N$ constraints
$p_{y_n}\approx 0$ and $\frac{1}{2}z_n^2\approx 0$ in the Cawley
counterexample. In terms of the total Hamiltonian $H_T$
\begin{equation}
 H_T=\sum_{n=1}^N{(p_{x_n}p_{z_n}-\frac{1}{2}y_nz_n^2)}+\sum_{n=1}^N{p_{y_n}}\nu_n,
 \end{equation}
 in Dirac-Bergmann formalism one has the secondary constraints
$\frac{1}{2}z_n^2\approx 0$, which are first-class in the Dirac
sense, and fourth-class in the Lusanna sense. The tertiary
constraints $p_{x_n}\approx 0$ given by Cawley are genuine
first-order equations of motion, whose phase space counterpart is
the Hamilton equations $p_{x_n}=\dot{z}_n=[z_n,H_T^F]\approx 0$,
where
$H_T^F\equiv\sum_{n=1}^N{p_{x_n}p_{z_n}}+\sum_{n=1}^N{\nu_np_{y_n}}$
is the final Dirac Hamiltonian ("$\equiv$" means strong equality in
the Dirac sense). In this system, $y_n$ is the gauge variable (gauge
degree of freedom\cite{29}), $z_n$ has the fixed value zero, while
$x_n$ is the physical degree of freedom. The problems of
interpretation in [28] are caused by the linearization of the
secondary constraints $z_n\approx 0$ instead of $\frac{1}{2}z_n^2\approx
0$. In the following, we will discuss this example again in our
procedure.

Adding $-\sum_{n=1}^N{\frac{d(\mu_n p_{y_n})}{dt}}$ determined by
$H_T$ (23) to the Lagrangian (17), we obtain a new Lagrangian
\begin{equation}
\begin{split}
L^1&=\sum_{n=1}^N{(\dot{x}_n\dot{z}_n+\frac{1}{2}y_nz_n^2-\dot{\mu}_np_{y_n}-\frac{1}{2}\mu_nz_n^2)}\\
&=\sum_{n=1}^N{(p_{x_n}p_{z_n}+\frac{1}{2}y_nz_n^2-\dot{\mu}_np_{y_n}-\frac{1}{2}\mu_nz_n^2)},
\end{split}
\end{equation}
and a new total Hamiltonian $H_T^1$
\begin{equation}
H^1_T=\sum_{n=1}^N{(p_{x_n}p_{z_n}-\frac{1}{2}y_nz_n^2+\nu^{\prime}_np_{y_n}+\frac{1}{2}\mu_nz_n^2)}
\end{equation}
with a new additional term $\sum_{n=1}^N{\frac{1}{2}\mu_nz_n^2}$,
where $\nu^{\prime}_n=\nu_n+\dot{\mu}_n$ is arbitrary, because
$\nu_n$ and $\mu_n$ are arbitrary functions of time. In this
stage, there appears new constraint $\frac{1}{2}z_n^2\approx 0$ in
r.h.s of (25). Considering the new constraint, and adding the
total time derivative
$-\sum_{n=1}^N\frac{d}{dt}(\frac{1}{2}\mu_nz_n^2)$ determined by
$H_T^1$ to the Lagrangian $L^1$, we obtain a new Lagrangian $L^2$
with the additional term $-\sum_{n=1}^N{\mu_nz_np_{x_n}}$, and its
corresponding Hamiltonian $H_T^2$ with the additional term
$\sum_{n=1}^N{\mu_nz_np_{x_n}}$. In the new stage, there appears
new constraint $z_np_{x_n}\approx 0$. Repeating a new annulation,
we obtain a new constraint $p_{x_n}^2\approx 0$ which is introduced
into the total Hamiltonian $H_T^2$ as a new $H_T^3$. Repeating a
new annulation again, we obtain nothing new and the procedure terminates.
For the Cawley counterexample, the extended Hamiltonian is
\begin{equation}
\begin{split}
H_E=&H_T^3=
\sum_{n=1}^N{(p_{x_n}p_{z_n}+\nu_n^{\prime}p_{y_n}+\xi_nz_n^2)}+\\
&+\sum_{n=1}^N{(\eta_nz_np_{x_n}+\mu_np_{x_n}^2)},
\end{split}
\end{equation}
and there are the first-class constraints
\begin{equation}
p_{y_n}\approx 0,\quad \frac{1}{2}z_n^2\approx 0, \quad
z_np_{x_n}\approx 0, \quad p_{x_n}^2\approx 0.
\end{equation}

In our procedure, there are more first-class constraints
$z_np_{x_n}\approx 0$ and $p_{x_n}^2\approx 0$ than that given by
Lusanna\cite{22}. The generator of the gauge transformations is
$G=\lambda_np_{y_n}+\mu_nz_n^2+\nu_nz_np_{x_n}+\omega_np_{x_n}^2$.
According to the gauge transformations determined by $G$, we can
obtain the Nother identities
\begin{equation}
L_{y_n}=0, \sqrt{2L_{y_n}}\frac{d}{dt}\sqrt{2L_{y_n}}=0,
(\frac{d}{dt}\sqrt{2L_{y_n}})^2=0,
\end{equation}
which demonstrates the point of view that the second Noether
theorem is the basis of the singular Lagrangians and Hamiltonian
constraints\cite{22}.

The Cawley constraints $z_n\approx 0$ and $p_{x_n}\approx 0$ can
be deduced from (27). $p_{x_n}\approx 0$ is the canonical equation
$p_{x_n}=\dot{z}_n=[z_n,H_E]\approx 0$. In terms of the
definitions of weak equality and strong equality, the constraints
$\frac{1}{2}z_n^2\approx 0$, $z_np_{x_n}\approx 0$ and
$p_{x_n}^2\approx 0$ are strong equations. They can be eliminated
from the extended Hamiltonian $H_E$ (28) which is eventually
simplified as the previous final Dirac Hamiltonian $H_T^F$
\cite{22}. The problem of the Cawley system is the linearization
of $\frac{1}{2}z_n^2\approx 0$, which means that the first-class
constraint $z_n\approx 0$ is substituted into the Poisson bracket
$[z_n^2,H_T]$. This process is in conflict with the definition of
weak equality in the Dirac sense.

In order to give the extremum of $S=\int{Ldt}$ corresponding to
the Lagrangian (17), the equations in (18) must include
$z_n\approx 0$ or $z_n^2\approx 0$, i.e., the secondary constraint. We
leave this check in the appendix. This result partly demonstrates
that under special conditions the extended Hamiltonian $H_E$ is better than the total
Hamiltonian $H_T$ in accordance with the original Lagrangian.

In the following we will consider a new example, in which one of
its Lagrange equations is a secondary constraint. It cannot be
generated by its total Hamiltonian, but it can be given by its
extended Hamiltonian. This system has the Lagrangian
\begin{equation}
L(x,y,z,\dot{x},\dot{y},\dot{z})=\dot{x}\dot{z}+\dot{y}e^x,
\end{equation}
and the Euler-Lagrange equations are
\begin{equation}
\ddot{x}=0,\quad e^x\dot{x}=0, \quad\ddot{z}-\dot{y}e^x=0.
\end{equation}
According to (29), the momenta $p_x$, $p_y$ and $p_z$ with respect
to $x$, $y$ and $z$ are
\begin{equation}
p_x=\dot{z},\quad p_y=e^x,\quad p_z=\dot{x},
\end{equation}
respectively, and then the primary constraint is
\begin{equation}
p_y-e^x\approx 0.
\end{equation}
The corresponding total Hamiltonian is
\begin{equation}
H_T=p_xp_z+\lambda(p_y-e^x),
\end{equation}
which generates the canonical equations
\begin{equation}
\begin{split}
& \dot{x}\approx p_z, \quad \dot{y}\approx \lambda,\quad
\dot{z}\approx p_x,\\
& \dot{p}_x\approx \lambda e^x,\quad \dot{p}_y\approx 0,\quad
\dot{p}_z\approx 0,
\end{split}
\end{equation}
where $\lambda$ is the Lagrange multiplier, an arbitrary function
of time. Comparing (34) with (30), it is obvious that the term
$e^x\dot{x}$ in (30) disappears in (34). The consistency of the
primary constraint (32) $\frac{d}{dt}(p_y-e^x)\approx 0$ can
generate a secondary constraint
\begin{equation}
e^xp_z=e^x\dot{x}\approx 0.
\end{equation}
Following our procedure, and adding the total time derivative
$-\frac{d}{dt}[\lambda(p_y-e^x)]$ determined by $H_T$ (33) to the
Lagrangian (29), one obtains a new total Hamiltonian
\begin{equation}
H^1_T=p_xp_z+\dot{\lambda}(p_y-e^x)+\lambda e^xp_z,
\end{equation}
and canonical equations
\begin{equation}
\begin{split}
& \dot{x}\approx p_z, \quad \dot{y}\approx \dot{\lambda},\quad
\dot{z}\approx p_x+\lambda e^x,\\
& \dot{p}_x\approx \dot{\lambda} e^x+\lambda e^xp_z,\quad
\dot{p}_y\approx 0,\quad \dot{p}_z\approx 0.
\end{split}
\end{equation}
It is easy to check that the canonical equations (37) and the two
constraints (32) and (35) completely contain the ones in (30). By
using $H^1_T$, we can easily prove that the two first-class
constraints (32) and (35) are two generators of gauge
transformations. The term $e^xp_z$ does not disappear in (37).
This result demonstrates that in accordance with the Lagrangian (29)
$H^1_T$ (36) is better than $H_T$
(33). The consistency of
(35) does not generate any new constraint, and the annulation
terminates. In this system, the extended Hamiltonian $H_E$ is the
$H^1_T$.

In summary, by adding the total time derivatives of all the
constraints to the Lagrangian under discussion step by step, we
achieve the further work left by Dirac, which is that the
first-class Poisson brackets $[\phi_m,H^{\prime}]$ of the
first-class Hamiltonian with an arbitrary primary constraint are
generators of gauge transformations. Hitherto, the Dirac
conjecture is eventually proved completely. All the first-class
constraints are generators of gauge transformations. In our
procedure, all higher-stage first-class constraints play the same
role as primary first-class constraints, and a complete set of
constraints are guaranteed to find because in the Poisson brackets
$[\chi_{m_i},\theta_{k^{\prime}_i}]$ and
$[\chi_{m_i},H^{\prime}_i]$, the $\chi_{m_i}$ denote all
constraints, irrespective of whether they are primary constraints
or higher-stage constraints, or whether they are first-class or
second-class. It is worth noting that for the Lagrangian the total
time derivatives are not meaningful, but they can be used to
reveal some new constraints from the original Lagrangian. This
conclusion may bring some new thoughts into classical theories.
The canonical equations given by $H_T$ and $H_E$ describe
equivalently the physical states except that obviously they show
different symmetries. In the procedure of quantization, we should
take $H_E$, because it considers more symmetries than $H_T$ under
special situations.

We would like to thank A. A. Deriglazov, Zi-Ping Li and Liu Zhao
for helpful discussions. This work is supported by the
National Natural Science Foundation of China
(under Grant 11047020, 11274166 and 11047173),
the National Basic Research Program of China (under Grant 2012CB921504)
and the Natural Science Foundation of Shandong
Province of China (under Grant ZR2012AM022, ZR2011AM019, ZR2010AQ025,
BS2010DS006, Y2008A14, and J08LI56).

\textbf{Appendix:} To formulate a variational problem, one needs
to specify two things: the action functional and boundary conditions.
We consider the system (17), whose action is
\begin{equation}
S=\int{dt(\dot{x}_n\dot{z}_n+\frac{1}{2}y_nz_n^2)}
\end{equation}
We take the functional (38) with boundary conditions
\begin{equation}
\vec{r}_n(0)\approx \vec{r}_0,\quad \vec{r}_n(T)\approx \vec{r}_T,
\end{equation}
and suppose $z_n(0)\neq z_n(T)$. This immediately gives the
equations in (19). Now we note that $z_n^2(t)\approx 0$ is in
contradiction with the supposition $z_n(0)\neq z_n(T)$. So, the
formulated problem has no solution: there is no trajectory which
starts at $\vec{r}_0$, terminates at $\vec{r}_T$ with $z_n(0)\neq
z_n(T)$, and gives (19).

Therefore, we need to start from the beginning. Let us take (38)
and the conditions (39) with the supposition $z_n(0)\approx
z_n(T)$. Then, evidently, $z_n^2(t)\approx 0$ or $z_n(t)\approx 0$
must be the equation of motion.

In resume, the variational problem defined by (38), (39) and
$z_n(0)\neq z_n(T)$ is not consistent. The problem defined by
(38), (39) and $z_n(0)\approx z_n(T)$ is consistent; its solution
is: $z_n(t)\approx 0$ and $x_n$, $y_n$ are arbitrary functions. In
other words, the extremum of (38) contains the secondary
constraints $z_n(t)\approx 0$ or $z_n^2(t)\approx 0$. For this
system, $H_E$ is better than $H_T$.

\end{document}